Molecular beam epitaxy growth of axion insulator candidate EuIn$_2$As$_2$

Muhsin Abdul Karim,[1,*] Jiashu Wang,[1,*] David Graf,[2] Kota Yoshimura,[1] Sara Bey,[1] Tatyana Orlova,[3] Maksym Zhukovskyi,[3] Xinyu Liu,[1] Badih A. Assaf[1]

[1] Department of Physics and Astronomy, University of Notre Dame, Notre Dame, IN, 46556

[2] National High Magnetic Fields Laboratory, Florida State University, Tallahassee, FL, 32310

[3] Notre Dame Integrated Imaging Facility, University of Notre Dame, Notre Dame, IN, 46556

[*] The two authors contributed equally to this manuscript.

**Abstract**. The synthesis of thin films of magnetic topological materials is necessary to achieve novel quantized Hall effects and electrodynamic responses. EuIn$_2$As$_2$ is a recently predicted topological axion insulator that has an antiferromagnetic ground state and an inverted band structure, but that has only been synthesized and studied as a single crystal. We report on the synthesis of c-axis oriented EuIn$_2$As$_2$ films on sapphire substrates by molecular beam epitaxy. By carefully tuning the substrate temperature during growth, we stabilize the Zintl phase of EuIn$_2$As$_2$ expected to be topologically non-trivial. The magnetic properties of these films reproduce those seen in single crystals, but their resistivity is enhanced when grown at lower temperatures. We additionally find that the magnetoresistance of EuIn$_2$As$_2$ is negative even up to fields as high as 31T. while it is highly anisotropic at low fields, it becomes nearly isotropic at high magnetic fields above 5T. Overall, the transport characteristics of EuIn$_2$As$_2$ appear similar to those of chalcogenide topological insulators, motivating the development of devices to gate tune the Fermi energy and reveal topological features in quantum transport.

I. **Introduction.**

Magnetic topological insulators with a ferromagnetic ground state are behind the recent discovery of the quantized Hall effect at zero magnetic field. [1] This discovery has since stimulated a search for topological materials that host complex magnetic ground states beyond ferromagnetism, which can lead to other interesting and technologically relevant topological phases. [2–5]

EuIn$_2$As$_2$ is a candidate material that falls in this category. It has a layered crystal structure consisting of alternating Eu and In$_2$As$_2$ planes stacked along its c-axis. [6] It exhibits antiferromagnetic order at low temperature resulting from the interaction between Eu atoms occupying neighboring layers. Our primary motivation to study this material comes from theoretical predictions that EuIn$_2$As$_2$ is an axion insulator with a zero Chern number and a quantized magnetoelectric coupling term. The axion insulator state results from a band inversion in a crystal with inversion symmetry that hosts antiferromagnetism that breaks time-reversal symmetry. [7] These predictions argue that EuIn$_2$As$_2$ is the first stoichiometric compound with intrinsic magnetic order to belong to this topological class. Magnetometry measurements of EuIn$_2$As$_2$ single crystal show the presence of an antiferromagnetic ordering at low temperatures with an in-plane magnetic easy axis. [8] More recently, neutron diffraction measurements reported observing a helical magnetic structure, challenging the prior belief that EuIn$_2$As$_2$ has a collinear magnetic ordering. [9] The topological axion phase predicted by ref. [7,9] is altered when the ground state is helimagnetic, but the topological character of the material was argued to be preserved. Experimental evidence of a band inversion in EuIn$_2$As$_2$ from angle resolved-photoemission spectroscopy (ARPES) measurements makes this material a very good candidate to search for novel physics stemming coexisting topological and magnetic order. [10,11]

The electrical transport properties of $EuIn_2As_2$ have also been studied in the past [6,8,12–15]. The material exhibits a peak in the resistivity at the Neel temperature followed by a drop at low temperature. $EuIn_2As_2$ also exhibits a negative magnetoresistance, common to many Eu-based antiferromagnets, including $EuB_6$, $EuIn_2P_2$ and (Eu,Gd)Se. [16–18] It is maximized at the Neel temperature but is maintained in the ordered state. [6]

All of the above studies on the structural, magnetic, and electrical properties of $EuIn_2As_2$ were conducted on single crystals. Without a doubt, there is a need for thin films of this material to probe magnetotransport signatures of predicted topological edge or surface states [9,19], as well as the magnetoelectric responses [20,21] potentially due to its quantized axion angle. Thin films would also enable the realization of gated Hall bars of this material needed to gate tune the Fermi energy, typically located in the valence band, into the bulk gap. [10] Without this, the native doping observed in previous studies would mask the contribution of topological edge or surface states. [10]$EuIn_2As_2$ cannot be mechanically exfoliated, so the development of its synthesis by molecular beam epitaxy (MBE) is needed. This is problem is challenging because the Zintl phase of $EuIn_2As_2$ competes with the highly studied and thermodynamically favorable zincblende arsenide phases at low substrate growth temperatures.

In this work, we overcome this challenge and successfully identify a temperature region in the MBE growth scheme at which the Zintl phase becomes thermodynamically favorable, yielding a layered $EuIn_2As_2$ structure. The onset of this phase occurs at a substrate temperature of 680°C, well above the ideal growth window of InAs. Using this developed growth scheme, we achieve $EuIn_2As_2$ films on (0001)-oriented $Al_2O_3$, which are close to 45-100nm thick, and reproduce the magnetic properties seen in bulk single crystals. The resistivity in our samples is likely controlled by film morphology as it is consistently larger than the resistivity in single crystals. We also observe a negative magnetoresistance consistent with the magnetic polaron picture, resulting from scattering due to magnetic fluctuations. Its anisotropy follows the magnetic anisotropy of the crystal at low magnetic fields but is suppressed at high magnetic fields despite remaining negative all the way up to 31T.

## II. Results
### A. Molecular beam epitaxy and characterization

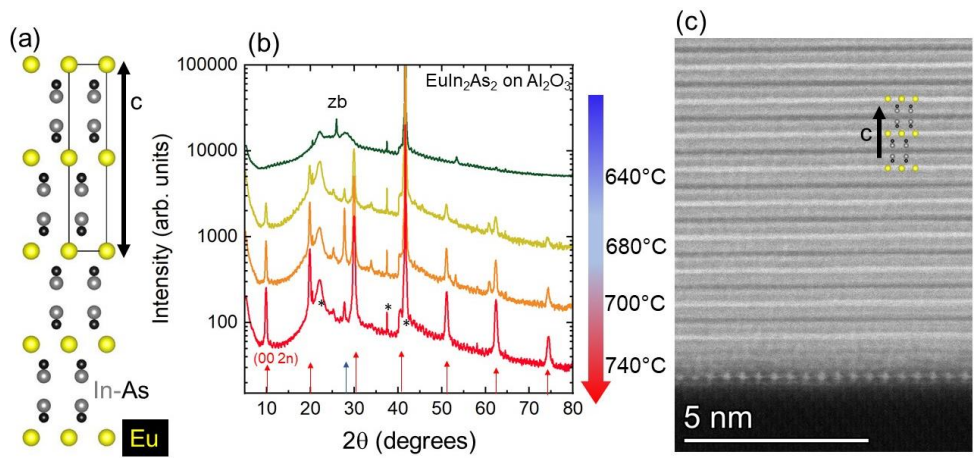

**FIG 1**. (a) Crystal structure of EuIn$_2$As$_2$. (b) X-ray diffraction patterns measured on a series of Eu-In-As films grown on sapphire between 640°C and 740°C. The (0 0 2n) Bragg series of EuIn$_2$As$_2$ is highlighted by red arrows. The (111) Bragg peak of the zincblende structure is labeled zb, and is seen clearly in the sample grown at 640°C. The * symbol marks the Bragg peaks of the sapphire substrate and instrument artifacts. The blue arrow denotes the position of the (102) Bragg peak of EuIn$_2$As$_2$. (c) High-angle annular dark-field scanning TEM image of a sample grown at 740°C. The films grown at 740°C and 680°C are 45nm thick, the one grown at 700°C is 75nm.

MBE of EuIn$_2$As$_2$ films is carried out on (0001) - oriented sapphire substrates. We chose sapphire (a=4.758Å) because of its ability to withstand high substrate temperatures, despite its lattice mismatch with EuIn$_2$As$_2$ (a=4.21Å). The temperatures of elemental sources are tuned to ensure beam equivalent pressures (BEPs) that maintain As-rich conditions through the growth. The As:Eu BEP ratio is between 20:1 and 29:1, while the As:In BEP ratio is kept close to 10:1. In, As (both 99.9999% pure) and Eu (99.99% pure) are evaporated from standard Knudsen-cells. The base pressure in the MBE is $5 \times 10^{-9} torr$. The substrate temperature T$_{sub}$ is varied between 640° and 740°C. As we shall show next, as T$_{sub}$ approaches the Indium cell temperature, exceeding that of Eu and As, the nucleation of elemental layers and zincblende phases is suppressed. Under these conditions, the Zintl phase is formed. Also, it is worth highlighting that T$_{sub}$ is significantly higher than what is commonly used for III-V zincblende materials. This requirement also motivates our substrate choice, since sapphire is known to remain structurally stable at these temperatures, unlike III-V semiconductor wafers. [22]

X-ray diffraction measurements are carried out using a Cu-K$\alpha$ source in the specular direction on films grown at various T$_{sub}$. They yield the patterns shown in Fig. 1(b). At $640°C$, the nucleation of InAs (possible Eu-doped) is thermodynamically favored and is evidenced by the peak highlighted in grey close to $2\theta = 25°$. As of $T_{sub} = 680°C$, a periodic pattern of peaks repeating almost every $10°$ emerges and grows stronger as temperature increases to $740°C$. These Bragg peaks are characteristic of the layered c-axis oriented Zintl phase of EuIn$_2$As$_2$ shown in Fig. 1(a) and agree with what has been previously reported. [6,8,9] From this pattern, we find a lattice constant c=17.874±0.003 Å (see analysis in Appendix A). It is 0.1% larger than what is reported by Zhang [8], 0.7% larger than what is found by Riberolles [9], but 0.08% smaller than what is found by Goforth [6] et al. in single crystals. It is within sample-to-sample variations seen in single crystals. For films grown at lower substrate temperatures, an additional Bragg peak matching the (102) EuIn$_2$As$_2$ Bragg line can be seen below 30°. It is however, dramatically weaker than the EuIn$_2$As$_2$ (006) for the film grown at $740°C$. Thus, with the appropriately tuned substrate temperature, we synthesize a dominantly c-axis oriented EuIn$_2$As$_2$ films on sapphire despite the large lattice mismatch between the in-plane lattice parameters of the two materials.

Cross-sectional transmission electron microscopy images were acquired using a double tilt holder and probe-corrected Spectra 30-300 transmission electron microscope (Thermo Fisher Scientific, USA) equipped with a field emission gun operated at 300 kV. The TEM image in Fig. 1(c) confirms the layered Zintl phase with a c-lattice constant close to 18Å, consistent with X-ray diffraction. An amorphous layer is seen near the interface, but its thickness does not exceed a single unit cell. Energy dispersive X-ray spectra acquired during these measurements yield the composition of the layer Eu$_{(1.05\pm0.10)}$In$_{(1.90\pm0.15)}$As$_{(2.05\pm0.10)}$.

### B. Magnetic Properties

Superconducting quantum interference device (SQUID) magnetometry measurements are performed next using a Quantum Design MPMS-5 system. They reveal properties expected for EuIn$_2$As$_2$ and are consistent with previous studies on single crystals. Fig. 2(a) shows temperature-dependent measurements of the magnetization of the 3 films grown above 680°C down to T=5K. A peak is systematically seen at 14K but is preceded by the onset magnetic order slightly below 20K, close to the Neel temperature reported in single crystals (17K). [6,8,15,23] The field dependent magnetization is also measured in all samples that contain the Zintl phase. We focus on the sample grown at 740°C, which shows the highest purity, but we note that the film grown at 700°C exhibits comparable properties. Fig. 2(b) plots the magnetization as a function of the magnetic field applied in the (ab)-plane of the film. It is evident from this plot that EuIn$_2$As$_2$ exits the antiferromagnetic ground state at low magnetic field. By 2T, the system is saturated to a ferromagnetic state with a saturation value close to 6.8 µ$_B$/Eu. We note that in this film, we also observe a small remanent magnetization of 0.3µ$_B$/Eu, likely due to an unpaired Eu layer at the surface, as in other A-type antiferromagnets. The first derivative of the M(H) curves is plotted in Fig. 2(c), highlighting a peak at 0.3T. It corresponds to the spin-flop transition of the system when the magnetic field is applied along the easy axis. In Fig. 2(d), we plot the magnetization measured with the field applied along two perpendicular directions, in-plane and out-of-plane, at 4K. The magnetization saturates faster in the in-plane direction, confirming in-plane anisotropy, consistent with a (001)-oriented EuIn$_2$As$_2$ film, with its c-axis along the growth direction.

Table 1 compares the magnetic properties of the sample grown at 740°C to those measured in single crystals. The saturation magnetization per Eu and the saturation field in the out-of-plane direction agree with previous findings. The spin-flop field extracted from dM/dH is found to be larger. However, the observed spin-flop peak in dM/dH is somewhat broad (±0.05T at half-maximum). The spin-flop transition in antiferromagnets is determined by the competition between the magnetic exchange and anisotropy energy, so that: [24–26]

$$H_{sf} = \sqrt{(H_A)(2H_E - H_A)}$$

The anisotropy ($H_A$) and exchange magnetic fields ($H_E$) can be determined by comparing the in-plane ($B_{sat}^{IP} = 1.3$) and out-of-plane ($B_{sat}^{OOP} = 2.3T$) saturation fields (Fig. 2(c)), as in [8]:

$$\mu_0 H_E = \frac{B_{sat}^{OOP} + B_{sat}^{IP}}{4} \text{ and } \mu_0 H_A = \frac{B_{sat}^{OOP} - B_{sat}^{IP}}{2}$$

We find $\mu_0 H_E \approx 0.9T$ and $\mu_0 H_A \approx 0.5T$. From that we can determine $\mu_0 H_{sf} \approx 0.8T$, which is inconsistent with experimental findings as also seen in [8]. This inconsistency implies that the spin-flop transition cannot be accounted for by a molecular field treatment that simply includes nearest neighbor exchange interactions and magnetic anisotropy. There must be competing exchange interactions influencing $H_{sf}$. Ref. [9] has found a broken helimagnetic state in EuIn$_2$As$_2$ at low temperature. But, so far, no work has reported a theoretical treatment computing the phase diagram of the material with the interactions responsible for the broken helix included. Those interactions, could account for the spin-flop transition occurring at such a low magnetic field. In our case, the broadening of the transition and the slight enhancement of the saturation fields, are due to the film morphology. This is discussed later in the manuscript.

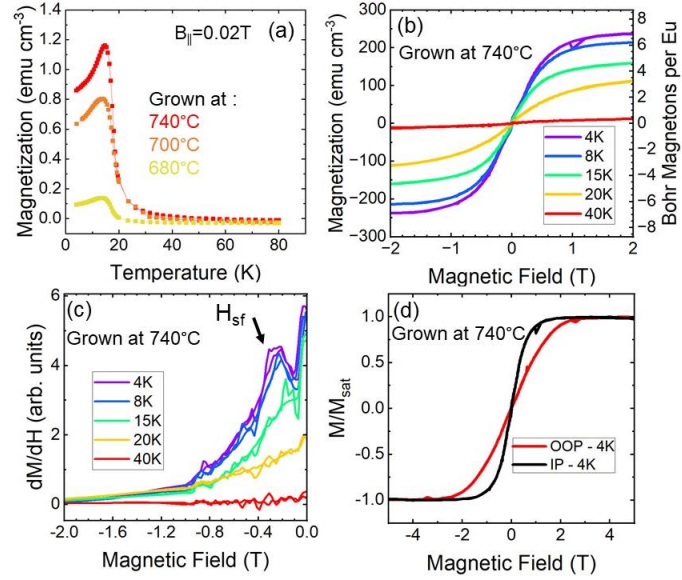

**FIG 2.** (a) Magnetization versus temperature for the three samples grown above 680°C. (b) Magnetization versus temperature for the sample grown at 740°C measured at B=0.02T applied in-plane. (c) Derivative of the magnetization with respect to magnetic field from the curves shown in (b). $H_{sf}$: spin-flop field. (d) Magnetization divided by its saturation plotted versus magnetic field applied out-of-plane (OOP) and in-plane (IP).

|  | $M_{sat}(\mu_B/Eu)$ | $B_{sat}^{OOP}$ (T) | $B_{sf}$ (T) |
|---|---|---|---|
| This work (t=45nm, 740°C) | 6.8 | 2.3 | 0.25±0.05 (4K) |
| [15] | 6.9 | 2 | 0.21 (2K) |
| [6] | 7.378 | 2 | - |
| [13] | 6.7 | 2 | 0.21 (10K) |
| [12] | 6.9 | 1.7 | - |
| [8] | 6.9 | 1.9 | 0.19 (2K) |

Table 1. Magnetic sample characteristics compared to those from singles crystals.

### C. Magnetotransport measurements

Magnetotransport measurements are carried out using a Quantum Design MPMS5 on the film exhibiting the Zintl phase and grown at 740°C. A magnetic field up to 7T is applied along the c-axis, and the measurements are carried out down to 4K. The measured sample is rectangular and is connected in a Hall configuration. The Hall resistance is first shown in Fig. 3(a) for different temperatures. It is robustly linear at high magnetic field above 2T but exhibits a non-linearity at low field. The charge carrier density extracted from the Hall slope is found to be $7.5\times10^{19}$ holes/cm$^3$, comparable with what is typically found in topological insulators such as uncompensated $Bi_2Te_3$ [27–29] and with previous work on single crystals of $EuIn_2As_2$. [10,12] Subtracting the slope of the normal Hall effect reveals an anomalous Hall component shown in Fig. 3(b). It does not exceed 0.55Ω at saturation. This value corresponds to a 2D Hall conductivity of $0.2e^2/h$ evidencing a strong anomalous Hall effect. Qualitatively, the anomalous Hall resistance has the same field dependence as the magnetization (Fig. 2(d)), both saturating at 2T when the field is applied along the c-axis. The Hall resistivity reaches 2.5μΩ.cm at saturation, comparable to what is found in single crystals. [12] In ref. [12], a non-monotonic Hall resistivity versus magnetic field, referred to as a topological Hall effect, is observed and is attributed to the non-coplanar spin texture of EuIn$_2$As$_2$. We do not see evidence of such a behavior in our films.

The temperature dependence of the resistivity is shown in Fig. 3(c). A prominent peak occurs at 18K consistent with the Neel temperature of EuIn$_2$As$_2$, and is followed by a drop in resistivity when the material enters its ordered magnetic ground state. This behavior is qualitatively consistent with what was seen in EuIn$_2$As$_2$ single crystals [6,12,15] and many other Eu-based compounds, regardless of topological character [17,18,30–32]. The drop is enhanced with increasing magnetic field, demonstrating an enhancement of the conductivity with increasing ferromagnetic saturation. The magnetoresistance of this sample is shown in Fig. 3(d). It is consistently negative up to 7T. This behavior was discussed in previous work on EuIn$_2$As$_2$ and EuIn$_2$P$_2$ and is thought to be a result of magnetic fluctuations that get suppressed by the applied magnetic field above the magnetic ordering temperature. [6,33,34] A similar behavior occurs when transport is dominated by hopping between magnetic polaron clusters and is suggested to happen in EuO, [30] (Eu,Gd)Se, [17] EuB$_6$, [18,31,32], manganites [35] and other Eu-based Zintl compounds. [33,36]

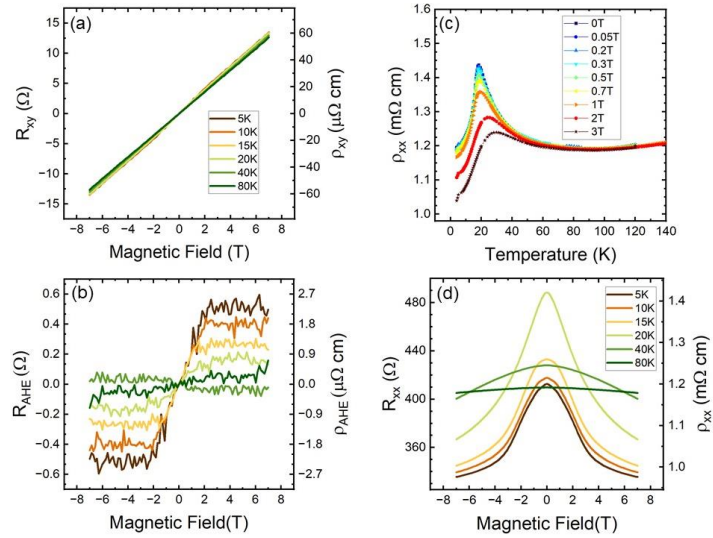

**FIG 3.** (a) Hall resistance R$_{xy}$ of the 45nm EuIn$_2$As$_2$ film grown at T$_{sub}$=740°C measured up to 7T between 5K and 80K. (b) Anomalous part of the Hall effect obtained for different temperatures after subtracting a linear slope from R$_{xy}$ at high magnetic field. (c) Resistivity of the same EuIn$_2$As$_2$ film versus temperature at different applied magnetic fields from 0 to 3T. (d) Resistance R$_{xx}$ as a function of applied magnetic field up to 7T.

To further understand this behavior, the MR is plotted as a function of magnetization relative to its saturation $m = M/M_{sat}$ in Fig. 4(a). The low field MR clearly varies as $MR = Cm^2$ consistent with the Majumdar-Littlewood relation. This ties the MR's behavior to the role of magnetic fluctuations and their suppression at high magnetic field. [35] From the Majumdar-Littlewood, we can relate $C$ to the charge carrier density and correlation length $\xi_0$ between fluctuating spins.

$$C = \left(\frac{1}{2}k_f \xi_0\right)^2 = \frac{1}{4}(3\pi^2 . n)^{2/3} \xi_0^2$$

Here, $k_f$ is the Fermi wavevector and $n = \frac{k_f^3}{3\pi^2}$ is the charge carrier density. The expression of $n$ in terms of $k_f$ assumes a spherical Fermi surface. The three-dimensional shape of the Fermi surface of EuIn$_2$As$_2$ is unknown, so our assumption only gives an effective $k_f$. In the diffusive regime, this choice of effective k$_f$ is more suitable than the assumption that $k_{||} = k_f$. [8] With this assumption, we recover a correlation length of 11Å at 20K, close to the out-of-plane lattice constant and to the separation between the Eu-planes.

The negative MR persists at low temperature, well below the Neel temperature. Its anisotropy further correlates with the behavior of the magnetization below the Neel temperature. We have measured its angular dependence up to 31T at 1.6K. This data is shown in Fig. 4(b). A large cusp is observed between ±5T. The cusp is broader when the field is applied along the c-axis (see inset of Fig. 4(b)), perpendicular to the magnetic easy axis of EuIn$_2$As$_2$. This confirms the MR saturates more slowly along the hard-axis and correlates it with the amount of the magnetic field that it takes to saturate the magnetization. This finding is consistent with the magnetic polaron picture even below the Neel temperature. At very high magnetic fields well above 2T and up to 31T, the MR remains negative, despite the magnetization saturating, but becomes nearly isotropic. In Fig. 4(c), we see evidence of this as the MR has a strong angular dependence at 1T and 2T but is nearly independent of angle at 5T and above. A likely contribution to the MR at very high magnetic fields could come from the spin-Zeeman and cyclotron energy altering the band dispersion of EuIn$_2$As$_2$. This band origin of the negative MR has been studied in the past, both in the extreme quantum limit and in the diffusive limit [37,38]. However, a good knowledge of the shape of the bulk Fermi surface of EuIn$_2$As$_2$ is required to conclusively tie the negative MR to a band origin.

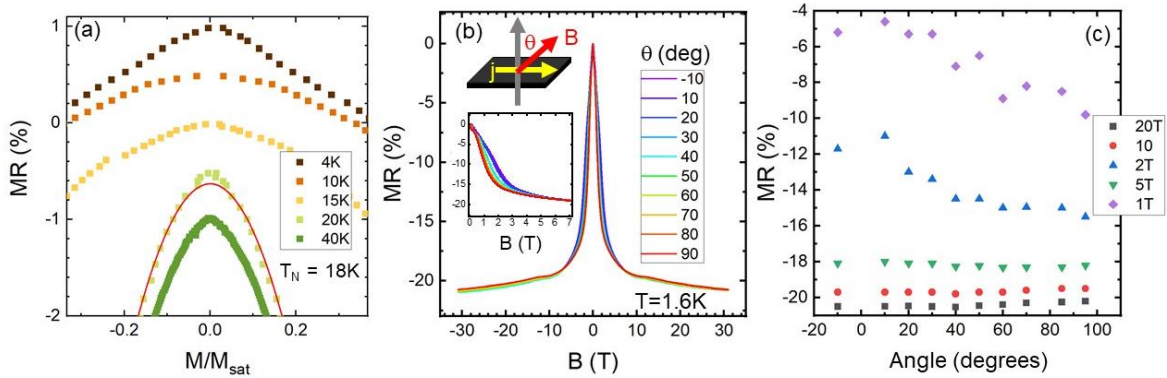

**FIG 4.** (a) Scaling of the MR with magnetization relative to its saturation M/M$_{sat}$ at different temperature. M$_{sat}$ is the saturation magnetization at 4K and 7T. The curves are shifted for clarity. The solid red lines are parabolic curve fits. (b) Angular dependence of the magnetoresistance up to 31T at T=1.6K. The insets show the direction of the magnetic field B with respect to the current and inset graph is a zoom in at low field (up to 7T). (c) Magnetoresistance plotted as a function of angle for different values of magnetic field.

The finding of negative MR originating from polaron hopping at low magnetic field is not unexpected for a Eu-containing semiconductor. However, the MR and its anisotropy are striking evidence that bulk states dominate magnetotransport signatures in this material. We must highlight that these bulk states clearly yield a magnetoresistance phenomenon vastly different from the weak antilocalization (WAL) behavior expected for topological surface states in the diffusive regime. We did not see any evidence of WAL in the films studied here. But, future work on thinner films, thin enough to quantum confine the bulk bands

(<20nm), can potentially settle whether topological surface states with spin-momentum locking [10] contribute to magnetotransport in EuIn$_2$As$_2$. We also note that the colossal size of the magnetoresistance (in magnetoconductance ($\Delta G_{xx} \gg e^2/h$)) also rules out weak localization as a possible origin.

### III. Discussion

The transport characteristics of our films are compared to those found in previous studies on single crystals in Table 2. The charge carrier density is consistent with what is typical of EuIn$_2$As$_2$ and is within sample-to-sample variations found in single crystals. The magnitude of the negative MR measured at 3T was smaller than what is reported in [6,13] but is also within sample-to-sample variations compared to prior work. The sample grown at 740°C has the highest mobility; it reaches 70cm$^2$/Vs. While this is too low to reach the Landau quantized regime at reasonable magnetic fields, it is consistent with previous work on single crystals (see table 2). The lowest resistivity was also measured in the sample grown at 740°C. It is 10 to 4 times larger than what was measured in single crystals.

Atomic force microscopy measurements shown in Fig. 5 shed light on discrepancies between films and single crystals. A surface roughness exceeding 10nm, as seen in Fig. 5, accounts for the enhanced resistivity reported in table 2. The sample grown at 680°C was not continuous, so its resistivity could not be measured. This indicates that a Stranski-Krastanov layer-plus-island process drives the nucleation of EuIn$_2$As$_2$ on lattice mismatched sapphire. Additionally, this morphology can explain the broadening of the spin-flop transition (Fig. 2(c)) and the slight enhancement of magnetic field phase boundaries (B$_{sf}$ and B$_{sat}$) compared to single crystals. The observed morphology can, in fact, lead to an inhomogeneous distribution of lattice and thermal strain, which can alter magnetic exchange interactions.

|  | Carrier density [cm$^{-3}$] | Resistivity [mΩ cm] | MR (%) at 3T, 10K |
|---|---|---|---|
| This work (t=45nm, 740°C) | p=7.5×10$^{19}$ | 1.19 | - 12.8% |
| [10] | p=6.5×10$^{19}$ | - | - |
| [13] | - | 0.4 | - 40% |
| [6] | - | 0.22 | - 50% |
| [39] | - | 0.32 | - |
| [12] | p=4.2×10$^{19}$ | 0.15 | - 10% |
| [8] | p=3.6x10$^{20}$ (from R$_H$) | 0.15 | + 5.6% |
| [40] | p=4.0×10$^{19}$ | 0.15 | - |
| [15] |  | 0.19 | - 10% (at 12K) |

Table 2. Magnetotransport characteristics compared to those from single crystals. MR=[R(B)-R(0)]/R(0).

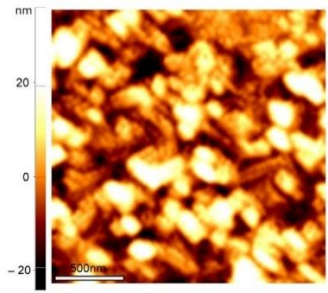

**FIG 5**. Atomic force microscopy image of a $2\mu m \times 2\mu m$ area of the film grown at 740°C.

### IV. Conclusions

We have thus successfully synthesized the Zintl phase of EuIn$_2$As$_2$ by MBE on sapphire. Our choice of substrate is primarily motivated by its ability to withstand a high substrate temperature during growth, which we found essential to stabilize the Zintl phase of EuIn$_2$As$_2$. The synthesis of EuIn$_2$As$_2$ on a lattice matched substrate should be developed next to enhance the mobility further and reduce layer roughness. The films grown on sapphire reproduce the magnetic properties of bulk single crystals but yield a broadened spin-flop transition. The realization of thin films enables the application of a gate voltage to tune the Fermi level of EuIn$_2$As$_2$. It also opens the door to magnetooptical measurements at long wavelengths in the infrared, THz and mm-wave parts of the spectrum. [41–43] Such measurements are needed to elucidate the magnetic field's impact on this material's band structure and to discover its predicted electrodynamic axion response.

**Appendix 1: Nelson Riley analysis of the X-ray diffraction pattern**

The Nelson -riley method allows us to determine the lattice parameter of EuIn$_2$As with improved precision. W first compute the c-lattice parameter using Bragg's law for each observed Bragg reflection ((00 2n) n=2-7). We then plot all the values of c that we find against the Nelson-Riley function [44]:

$$f(\theta) = \frac{1}{2}[(\cos^2\theta)/\sin(\theta) + (\cos^2\theta)/\theta]$$

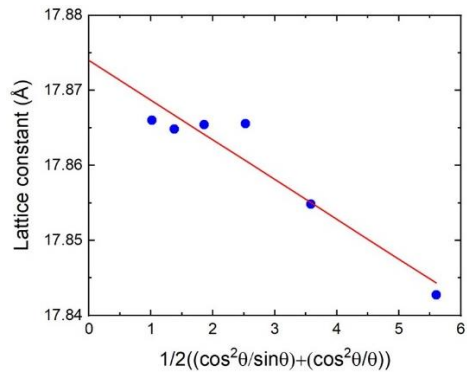

**FIG 6.** Nelson-Riley analysis for the sample grown at 740°C.

The plot is shown in Fig. 6. It yields c=17.874±0.003 at the intercept between a linear fit to the data and the y-axis. This point represents the point for which various sources of uncertainty are minimized.

Acknowledgements. Work supported by NSF-DMR-1905277. A portion of this work was performed at the National High Magnetic Field Laboratory, which is supported by National Science Foundation Cooperative Agreement No. DMR-2128556 and the State of Florida.

[1]  C.-Z. Chang, J. Zhang, X. Feng, J. Shen, Z. Zhang, M. Guo, K. Li, Y. Ou, P. Wei, L.-L. Wang, Z.-Q. Ji, Y. Feng, S. Ji, X. Chen, J. Jia, X. Dai, Z. Fang, S.-C. Zhang, K. He, Y. Wang, L. Lu, X.-C. Ma, and Q.-K. Xue, *Experimental Observation of the Quantum Anomalous Hall Effect in a Magnetic Topological Insulator.*, Science (1979) **340**, 167 (2013).

[2]  Y. Tokura, K. Yasuda, and A. Tsukazaki, *Magnetic Topological Insulators*, Nature Reviews Physics **1**, 126 (2019).

[3]     N. P. Armitage, E. J. Mele, and A. Vishwanath, *Weyl and Dirac Semimetals in Three-Dimensional Solids*, Rev Mod Phys **90**, 015001 (2018).

[4]     C.-Z. Chang, C.-X. Liu, and A. H. MacDonald, *Colloquium : Quantum Anomalous Hall Effect*, Rev Mod Phys **95**, (2023).

[5]     C.-Z. Chang, P. Wei, and J. S. Moodera, *Breaking Time Reversal Symmetry in Topological Insulators*, MRS Bull **39**, 867 (2014).

[6]     A. M. Goforth, P. Klavins, J. C. Fettinger, and S. M. Kauzlarich, *Magnetic Properties and Negative Colossal Magnetoresistance of the Rare Earth Zintl Phase EuIn2As2*, Inorg Chem **47**, 11048 (2008).

[7]     Y. Xu, Z. Song, Z. Wang, H. Weng, and X. Dai, *Higher-Order Topology of the Axion Insulator EuIn2As2*, Phys Rev Lett **122**, 256402 (2019).

[8]     Y. Zhang, K. Deng, X. Zhang, M. Wang, Y. Wang, C. Liu, J.-W. Mei, S. Kumar, E. F. Schwier, K. Shimada, C. Chen, and B. Shen, *In-Plane Antiferromagnetic Moments and Magnetic Polaron in the Axion Topological Insulator Candidate EuIn2As2*, Phys Rev B **101**, 205126 (2020).

[9]     S. X. M. Riberolles, T. V. Trevisan, B. Kuthanazhi, T. W. Heitmann, F. Ye, D. C. Johnston, S. L. Bud'ko, D. H. Ryan, P. C. Canfield, A. Kreyssig, A. Vishwanath, R. J. McQueeney, L. L. Wang, P. P. Orth, and B. G. Ueland, *Magnetic Crystalline-Symmetry-Protected Axion Electrodynamics and Field-Tunable Unpinned Dirac Cones in EuIn2As2*, Nat Commun **12**, 999 (2021).

[10]    T. Sato, Z. Wang, D. Takane, S. Souma, C. Cui, Y. Li, K. Nakayama, T. Kawakami, Y. Kubota, C. Cacho, T. K. Kim, A. Arab, V. N. Strocov, Y. Yao, and T. Takahashi, *Signature of Band Inversion in the Antiferromagnetic Phase of Axion Insulator Candidate EuIn2As2*, Phys Rev Res **2**, 33342 (2020).

[11]    S. Regmi, M. M. Hosen, B. Ghosh, B. Singh, G. Dhakal, C. Sims, B. Wang, F. Kabir, K. Dimitri, Y. Liu, A. Agarwal, H. Lin, D. Kaczorowski, A. Bansil, and M. Neupane, *Temperature-Dependent Electronic Structure in a Higher-Order Topological Insulator Candidate EuIn2As2*, Phys Rev B **102**, 1 (2020).

[12]    J. Yan, Z. Z. Jiang, R. C. Xiao, W. J. Lu, W. H. Song, X. B. Zhu, X. Luo, Y. P. Sun, and M. Yamashita, *Field-Induced Topological Hall Effect in Antiferromagnetic Axion Insulator Candidate EuIn2As2*, Phys Rev Res **4**, 013163 (2022).

[13]    F. H. Yu, H. M. Mu, W. Z. Zhuo, Z. Y. Wang, Z. F. Wang, J. J. Ying, and X. H. Chen, *Elevating the Magnetic Exchange Coupling in the Compressed Antiferromagnetic Axion Insulator Candidate EuIn2As2*, Phys Rev B **102**, 180404 (2020).

[14]    S. M. Kauzlarich, A. Zevalkink, E. Toberer, and G. J. Snyder, *Zintl Phases: Recent Developments in Thermoelectrics and Future Outlook*, in *Thermoelectric Materials and Devices* (The Royal Society of Chemistry, 2016), pp. 1–26.

[15]    T. Tolinski and D. Kaczorowski, *Magnetic Properties of the Putative Higher-Order Topological EuIn2As2*, SciPost Phys. Proc.: Proceedings of the International Conference on Strongly Correlated Electron Systems, Amsterdam **11**, 005 (2022).

[16] J. Jiang and S. M. Kauzlarich, *Colossal Magnetoresistance in a Rare Earth Zintl Compound with a New Structure Type: EuIn 2P 2*, Chemistry of Materials **18**, 435 (2006).

[17] S. Von Molnar and S. Methfessel, *Giant Negative Magnetoresistance in Ferromagnetic Eu1-XGd XSe*, J Appl Phys **38**, 959 (1967).

[18] S. Von Molnar, J. M. Tarascon, and J. Etourneau, *Transport and Magnetic Properties of Carbon Doped EuB6*, J Appl Phys **52**, 2158 (1981).

[19] Y. Xu, Z. Song, Z. Wang, H. Weng, and X. Dai, *Higher-Order Topology of the Axion Insulator EuIn2As2*, Phys Rev Lett **122**, 256402 (2019).

[20] A. Essin, J. Moore, and D. Vanderbilt, *Magnetoelectric Polarizability and Axion Electrodynamics in Crystalline Insulators*, Phys Rev Lett **102**, 146805 (2009).

[21] J. Ahn, S. Y. Xu, and A. Vishwanath, *Theory of Optical Axion Electrodynamics and Application to the Kerr Effect in Topological Antiferromagnets*, Nat Commun **13**, 7615 (2022).

[22] M. Yano, M. Nogami, Y. Matsushima, and M. Kimata, *Molecular Beam Epitaxial Growth of InAs*, Jpn J Appl Phys **16**, 2131 (1977).

[23] P. F. S. Rosa, C. Adriano, T. M. Garitezi, R. A. Ribeiro, Z. Fisk, and P. G. Pagliuso, *Electron Spin Resonance of the Intermetallic Antiferromagnet EuIn2As2*, Phys Rev B **86**, 094408 (2012).

[24] S. Foner, *High-Field Antiferromagnetic Resonance in Cr2O3*, Physical Review **130**, 183 (1963).

[25] I. S. Jacobs, *Spin-Flopping in MnF2 by High Magnetic Fields*, J Appl Phys **32**, S61 (1961).

[26] J. Coey, *Magnetism and Magnetic Materials* (Cambridge University Press, 2010).

[27] O. Caha, a Dubroka, V. Holy, H. Steiner, O. Rader, T. N. Stanislavchuk, a a Sirenko, G. Bauer, and G. Springholz, *Growth, Structure, and Electronic Properties of Epitaxial Bismuth Telluride Topological Insulator Films on BaF 2 (111) Substrates*, Cryst Growth Des **2**, 3365 (2013).

[28] N. Peranio, M. Winkler, M. Dürrschnabel, J. König, and O. Eibl, *Assessing Antisite Defect and Impurity Concentrations in Bi 2Te3 Based Thin Films by High-Accuracy Chemical Analysis*, Adv Funct Mater **23**, 4969 (2013).

[29] B. A. Assaf, T. Cardinal, P. Wei, F. Katmis, J. S. Moodera, and D. Heiman, *Linear Magnetoresistance in Topological Insulator Thin Films: Quantum Phase Coherence Effects at High Temperatures*, Appl Phys Lett **102**, 012102 (2013).

[30] J. B. Torrance, M. W. Shafer, and T. R. McGuire, *Bound Magnetic Polarons and the Insulator-Metal Transition in EuO*, Phys Rev Lett **29**, 1168 (1972).

[31] M. Pohlit, S. Rößler, Y. Ohno, H. Ohno, S. von Molnár, Z. Fisk, J. Müller, and S. Wirth, *Evidence for Ferromagnetic Clusters in the Colossal-Magnetoresistance Material EuB*, Phys Rev Lett **120**, 257201 (2018).

[32] G. Weill, I. A. Smirnov, and V. N. Gurin, *PRESSURE MEASUREMENTS OF ELECTRICAL TRANSPORT PROPERTIES OF EuB 6*, Le Journal de Physique Colloques **41**, C5 185 (1980).

[33] A. M. Goforth, H. Hope, C. L. Condron, S. M. Kauzlarich, N. Jensen, P. Klavins, S. MaQuilon, and Z. Fisk, *Magnetism and Negative Magnetoresistance of Two Magnetically Ordering, Rare-Earth-Containing Zintl Phases with a New Structure Type: EuGa 2Pn2 (Pn=P, As)*, Chemistry of Materials **21**, 4480 (2009).

[34] F. Pfuner, L. Degiorgi, H.-R. Ott, A. Bianchi, and Z. Fisk, *Magneto-Optical Behavior of EuIn2P2*, Phys Rev B **77**, 024417 (2008).

[35] P. Majumdar and P. B. Littlewood, *Dependence of Magnetoresistivity on Charge-Carrier Density in Metallic Ferromagnets and Doped Magnetic Semiconductors*, Nature **395**, 479 (1998).

[36] P. Rosa, Y. Xu, M. Rahn, J. Souza, S. Kushwaha, L. Veiga, A. Bombardi, S. Thomas, M. Janoschek, E. Bauer, M. Chan, Z. Wang, J. Thompson, N. Harrison, P. Pagliuso, A. Bernevig, and F. Ronning, *Colossal Magnetoresistance in a Nonsymmorphic Antiferromagnetic Insulator*, NPJ Quantum Mater **5**, 52 (2020).

[37] B. A. Assaf, T. Phuphachong, E. Kampert, V. V. Volobuev, P. S. Mandal, J. Sánchez-Barriga, O. Rader, G. Bauer, G. Springholz, L. A. De Vaulchier, and Y. Guldner, *Negative Longitudinal Magnetoresistance from the Anamolous N=0 Landau Level in Topological Materials*, Phys Rev Lett **119**, 106602 (2017).

[38] X. Dai, Z. Z. Du, and H.-Z. Lu, *Negative Magnetoresistance without Chiral Anomaly in Topological Insulators*, Phys Rev Lett **119**, 166601 (2017).

[39] B. Xu, P. Marsik, S. Sarkar, F. Lyzwa, Y. Zhang, B. Shen, and C. Bernhard, *Infrared Study of the Interplay of Charge, Spin, and Lattice Excitations in the Magnetic Topological Insulator EuIn2As2*, Phys Rev B **103**, 245101 (2021).

[40] K. Shinozaki, Y. Goto, K. Hoshi, R. Kiyama, N. Nakamura, A. Miura, C. Moriyoshi, Y. Kuroiwa, H. Usui, and Y. Mizuguchi, *Thermoelectric Properties of the as/p-Based Zintl Compounds Euin2as2-Xpx(X=0-2) and Srsn2as2*, ACS Appl Energy Mater **4**, 5155 (2021).

[41] L. Wu, M. Salehi, N. Koirala, J. Moon, S. Oh, and N. P. Armitage, *Quantized Faraday and Kerr Rotation and Axion Electrodynamics of a 3D Topological Insulator*, Science (1979) **354**, 1124 (2016).

[42] J. Wang, X. Liu, T. Wang, M. Ozerov, and B. A. Assaf, *G Factor of Topological Interface States in Pb1-XSnxSe Quantum Wells*, Phys Rev B **107**, 155307 (2023).

[43] J. Wang, T. Wang, M. Ozerov, Z. Zhang, J. Bermejo-Ortiz, S.-K. Bac, H. Trinh, M. Zhukovskyi, T. Orlova, H. Ambaye, J. Keum, L.-A. de Vaulchier, Y. Guldner, D. Smirnov, V. Lauter, X. Liu, and B. A. Assaf, *Energy Gap of Topological Surface States in Proximity to a Magnetic Insulator*, Commun Phys **6**, 200 (2023).

[44] J. B. Nelson and D. P. Riley, *An Experimental Investigation of Extrapolation Methods in the Derivation of Accurate Unit-Cell Dimensions of Crystals*, Proceedings of the Physical Society **57**, 160 (1945).